\documentclass[twocolumn,usenatbib]{mnras}
\usepackage{graphicx}
\usepackage{epsfig}
\usepackage{psfig}

\usepackage{amssymb}
\usepackage{natbib}

\newcommand{\lya}{Ly~$\alpha$ }

\def\CIV{C\,{\sc iv }}
\def\CIVab{C\,{\sc iv}$\lambda\lambda$1548,1551}
\def\CIVwave{C\,{\sc iv}$\lambda$1549}
\def\CIII{C\,{\sc iii}]}

\def\SiIIa{Si\,{\sc ii}$\lambda$1304}

\def\SiIIIa{Si\,{\sc iii}$\lambda$1206}

\def\SiIVab{Si\,{\sc iv}$\lambda\lambda$1393,1402}

\def\MgII{Mg\,{\sc ii}}
\def\MgIIwave{Mg\,{\sc ii}$\lambda$2798}

\def\MgIIab{Mg\,{\sc ii}$\lambda\lambda$2798,2803}

\def\OI{[O\,{\sc i}]}
\def\OIa{[O\,{\sc i}]$\lambda$1302}

\def\OIIIb{[O\,{\sc iii}]$\lambda$5007}

\def\OVIab{O\,{\sc vi}$\lambda\lambda$1031,1037}

\def\NVab{N\,{\sc v}$\lambda\lambda$1239,1242}

\def\CaIIab {Ca\,{\sc ii}$\lambda\lambda$3934,3969}

\def\zabs{$z_{\rm abs}$}
\def\zem{$z_{\rm em}$}
\def\Wr{$W_{\rm r}$}

\def\kms{$\rm km\,s^{-1}$}
\def\ergs{${\rm erg\,s^{-1}}$}
\def\sig{$\rm \sigma$ }

\title[Narrow \CIV\ absorption doublets]{Narrow \CIV\ absorption doublets on quasar spectra of the Baryon Oscillation Spectroscopic Survey}
\author[Zhi-Fu Chen et al.]{Zhi-Fu Chen$^{1, 2, 3, 4}$\thanks{E-mail:zhichenfu@126.com}, Qiu-Sheng Gu$^{1, 3, 4,}$, Luwenjia Zhou$^{1, 3, 4}$, and Yan-Mei Chen$^{1, 3, 4}$\\
$^{1}$ School of Astronomy and Space Science, Nanjing University, Nanjing 210093, P. R. China \\
$^{2}$ Department of Physics and Telecommunication Engineering, Baise University, Baise 533000, China\\
$^{3}$ Key Laboratory of Modern Astronomy and Astrophysics, Nanjing University, Nanjing 210093, P. R. China \\
$^{4}$ Collaborative Innovation Center of Modern Astronomy and Space Exploration, Nanjing, 210093, P.R. China\\}
\begin{document}
\maketitle

\begin{abstract}
In this paper, we extend our works of Papers I and II \citep[][]{2014ApJS..210....7C,2014ApJS..215...12C}, which are assigned to systematically survey \CIVab\ narrow absorption lines (NALs) with \zabs$\ll$\zem\ on quasar spectra of the Baryon Oscillation Spectroscopic Survey (BOSS), to collect \CIV\ NALs with \zabs$\approx$\zem\ from blue to red wings of \CIVwave\ emission lines. Together with Papers I and II, we have collected a total number of 41,479 \CIV\ NALs with $1.4544\le$\zabs$\le4.9224$ in surveyed spectral region redward of \lya\ until red wing of \CIVwave\ emission line. We find that the stronger \CIV\ NALs tend to be the more saturated absorptions, and associated systems (\zabs$\approx$\zem) seem to have larger absorption strengths when compared to intervening ones (\zabs$\ll$\zem). The redshift density evolution behavior of absorbers (the number of absorbers per redshift path) is similar to the history of the cosmic star formation. When compared to the quasar-frame velocity ($\beta$) distribution of \MgII\ absorbers, the $\beta$ distribution of \CIV\ absorbers is broader at $\beta\approx0$, shows longer extended tail, and exhibits a larger dispersion for environmental absorptions. In addition, for associated \CIV\ absorbers, we find that low-luminosity quasars seem to exhibit smaller $\beta$ and stronger absorptions when compared to high-luminosity quasars.
\end{abstract}


\section{Introduction}
\label{sect1}
The circumgalactic medium (CGM) surrounding galaxies is crucial to comprehending the feedback from and gas accretion into central objects, as well as galaxy evolution. In recent years, there are a number of successful results of directly detecting emission lines arising from CGM by deep imaging or spectroscopy observations \cite[][]{2014ApJ...786..106M,2013ApJ...766...58H,2014Natur.506...63C,2015MNRAS.452.2388H,2015Sci...348..779H,2015Natur.524..192M,2016arXiv160402942A}. However, it is still difficult to systematically survey the emission of CGM due to the low gas density. Luckily, the CGM would leave marks (so called absorption features) on the spectra of the background objects, which is an efficient way to investigate the properties and kinematics of CGM.

The gas, which infalls from the intergalatic medium (IGM) into galaxies to fuel star formation and is evaporated from galaxies to enrich the metal abundance of IGM, can be well characterized by metal absorption lines \citep[e.g.,]{2003MNRAS.339..289S,2009Natur.457..451D,2012ApJ...750..165R,2012MNRAS.421...98D,2012ApJ...747L..26R,2013Sci...341...50B,2014MNRAS.444.1260F,
2015ApJ...813...46B,2015ApJ...815...22K,2015ApJ...812...83N,2015ApJ...811..132M}.
The cold gas with $T\rm \sim 10^4$ --- $\rm 10^5~K$ of CGM has been explored by various transitions from neutral hydrogen to high-ionization absorption lines, such as the absorptions of \lya\ \citep[e.g.,][]{2006ApJ...638...52K,2014MNRAS.440..225B}, \CaIIab\ \citep[e.g.,][]{2013ApJ...773...16Z,2015MNRAS.452..511M}, \MgIIab\ \citep[e.g.,][]{2011ApJ...733..105K,2014MNRAS.441..886F,2016MNRAS.455.1713H}, \SiIIIa\ \citep[e.g.,][]{2013ApJ...765...27K}, \SiIVab\ \citep[e.g.,][]{2005AJ....130.1996S,2011ApJ...729...87C}, \CIVab\ \citep[e.g.,][]{2011ApJ...731...11C,2013ApJ...779L..17B}, \NVab\ \citep[e.g.,][]{2007A&A...465..171F,2009A&A...503..731F,2009A&A...496...31F}, and  \OVIab\ \citep[e.g.,][]{2008ApJS..177...39T,2011Sci...334..952T,2014ApJS..212....8S}. The properties of absorbers (e.g., ionization state, strength, kinematics) are related to halo mass of galaxy, locations and distances between the center of galaxy and absorber, colors, types, luminosities, morphologies, inclination angles, and star formation rates \citep[e.g.,][]{2007ApJ...662..909K,2015ApJ...815...22K,2011MNRAS.418.2730G,2014MNRAS.444.1260F,2014MNRAS.441..886F,2015ApJ...811..132M}. Thus, absorption lines provide key insight into galaxy's CGMs, feedback from central objects, evolutions as well as intrinsic properties of host galaxies.

Outflows and/or jets are fundamental components of active galactic nucleus (AGNs), which play a role in the galaxy evolution by the form of feedback. They carry angular momentum, energy and huge amounts of material away from production sites into host galaxies, CGMs and even IGMs to regulate their star formation rate, metallicities, and element distributions. The outflows are often detected via blueshifted absorption lines. Broad absorption lines (BALs) imprinted on quasar spectra are undoubtedly intrinsic to quasar systems and are likely to be formed in quasar outflows. Narrow absorption lines (NALs) with line widths of a few hundreds \kms\ are very common on quasar spectra. Some previous studies \citep[e.g.,][]{2007ApJ...669...74L,2008ApJ...679..239V,2012ApJ...748..131S,2013JApA...34..357P} have revealed that a part of NALs with \zabs $\rm \approx$ \zem\ is also intrinsic to quasar systems. Variation analysis of NALs suggests that some of NALs are probably produced in quasar outflows  \citep[e.g.,][]{2004ApJ...613..129W,2004ApJ...601..715N,2009NewAR..53..128C,2011MNRAS.410.1957H,2012ASPC..460...37C,2013MNRAS.434.3275C,2015MNRAS.450.3904C}.
And also, NALs are more common than BALs on quasar spectra especially for \MgII,  because i) low ionization BALs (LoBALs) are rare, and ii) a significant fraction of \MgII\ NALs are formed in intervening absorbers (i.e., not physically related to quasar themselves). Therefore, taking full advantage of NALs is crucial to understanding quasar outflows and their influences on surrounding environment.

The Sloan Digital Sky Survey \citep[SDSS,][]{2000AJ....120.1579Y} is a widely ambitious and influential survey in the astronomical history, which used a dedicated 2.5-meter telescope \citep[][]{2006AJ....131.2332G} at Apache Point Observatory, New Mexico, to obtain optical spectra with coverage from $\rm \lambda = 3800$ --- 9200\AA\ at a resolution of $\rm R \approx 2000$ \citep[][]{2009ApJS..182..543A} during the first (SDSS-I, 2000-2005) and second (SDSS-II, 2005-2008) stages. Some groups \citep[e.g.,][]{2005ApJ...628..637N,2011AJ....141..137Q,2013ApJ...770..130Z,2013ApJ...763...37C,2013ApJ...779..161S} have systematically survey NALs on quasar spectra of the SDSS-I/II. However, most of them only focused on intervening NALs (\zabs $\ll$ \zem), which limits the utilization of NALs to study quasar outflows and their influence on surrounding environment.

The SDSS program continued with the Third Sloan Digital Sky Survey (SDSS-III) using the same 2.5-meter telescope with update spectrographs \citep[][]{2013AJ....146...32S}, which collected data from July 2008 to June 2014 and obtained spectra with wavelength range from 3600\AA\ to 10400\AA\ at a resolution of $\rm R = 1300$ --- 2500 \citep[][]{2011AJ....142...72E}. The Baryon Oscillation Spectroscopic Survey \citep[BOSS,][]{2013AJ....145...10D} is the main dark time of legacy survey of the SDSS-III. The first spectral data release of the BOSS (SDSS DR9) includes 87,822 quasars \citep[DR9Q,][]{2012A&A...548A..66P}. On quasar spectra of the DR9Q, our early systematical surveys \cite[][here after Papers I and II]{2014ApJS..210....7C,2014ApJS..215...12C} of \CIV\ NALs are focused on intervening NALs, which are similar to the surveys mentioned in the last paragraph. In order to systematically collect \CIV\ NALs with \zabs$\rm\approx$\zem, we extend our program of Papers I and II to spectral data redward of \CIVwave\ emission lines. These NALs will favour future studies on AGN feedback and its influence on central objects and surrounding environment.

The paper is organized as follows. Section \ref{sect2} describes quasar sample selection, definition of surveyed spectral region, spectral analysis and NAL selection. We calculate the redshift path and estimate the completeness of our absorption line detection in Section \ref{sect3}. The strength and saturability of absorptions are characterized in Section \ref{sect4}. Section \ref{sect5} is discussions, which statistically describes redshift density evolution of absorbers, $\beta$ (see Equation \ref{eq7} for the definition of $\beta$) distributions, dependence of $\beta$ and absorption strengths on radio detection and UV continuum luminosity. A short summary is presented in Section \ref{sect6}. Throughout this paper, we adopt cosmological parameters of $\rm \Omega_M=0.3$, $\rm \Omega_\Lambda=0.7$, and $\rm H_0=70~km~s^{-1}~Mpc^{-1}$.

\section{Quasar sample and spectral analysis}
\label{sect2}
\subsection{Quasar sample and surveyed spectral region}
In this paper, we extend our previous search of \CIV\ NALs released in Papers I and II to spectral data redward of \CIVwave\ emission lines. Therefore, the quasar sample and selection criteria is the combination of Papers I and II. We briefly summarize the criteria used to construct quasar sample and surveyed spectral region as follows:

\begin{enumerate}
  \item Surveyed spectral region. Noisy data shortward of 3800\AA\ at observed frame, and the contamination of series of absorptions, for example, doublets of \OIa\ and \SiIIa, and \lya\ forest, lead to difficulties in searching for \CIV\ NALs. Thus, spectral region is limited to the data longward of the maximum value of 3800\AA\ and $\rm 1310\times$(1+\zem)\AA\  at observed frame. Papers I and II mainly paid attention to intervening NALs. Therefore, another limit is adopted to constrain the data blueward of 10,000 \kms\ of \CIVwave\ emission lines.
  \item Signal-to-noise ratio per pixel ($\rm \langle S/N \rangle$). The data accompanied with large noisy would markedly reduce the surveyed efficiency of NALs and may greatly lead to misidentification. In surveyed spectral region, the $\rm \langle S/N \rangle$ of all quasar spectra in the DR9Q has a median value of about $4~pixel^{\rm -1}$. The cut $\rm \langle S/N \rangle \ge 4$ $pixel^{\rm -1}$ is accepted to guarantee robust absorption line measurements.
\end{enumerate}

\begin{figure}
\centering
\includegraphics[width=7.5cm,height=5.8cm]{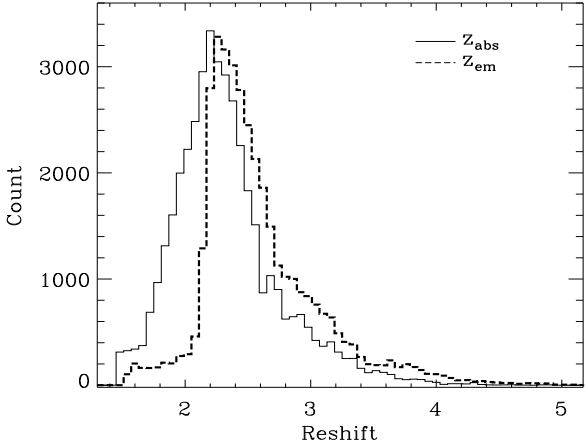}
\caption{Redshisft distributions. The black dash line is the \zem\ of quasars used to search for \CIV\ NALs, and the black solid one is the \zabs\ of all \CIV\ NALs from the current paper as well as Papers I and II}.
\label{fig1}
\end{figure}

In terms of above limits, Papers I and II selected 37,241 quasars with $\rm 1.54 \lesssim$ \zem $\rm \lesssim 5.16$ from the DR9Q that contains 87,822 quasars. Using the spectra of these 37,241 quasars whose \zem\ are displayed with black dash line in Figure \ref{fig1}, in this paper we survey \CIV\ NALs with \zabs\ $\rm \approx$ \zem\ in spectral data from blueward of 10,000 \kms\ until redward of 20,000 \kms\ of \CIVwave\ emission lines.

\subsection{Spectral analysis}

\cite{2011AJ....141..137Q} automatically searched for \MgII\ doublets on quasar spectra of the SDSS DR4. Using the same parent sample of \cite{2011AJ....141..137Q},  \cite{2013PASJ...65....8Q} visually inspected \MgII\ doublets one by one, and detected much more \MgII\ doublets. Although the visual detection requires a huge amount of time, we follow the same method of \cite{2013PASJ...65....8Q}, which is also used in Papers I and II, to survey \CIV\ NALs with \zabs $\rm \approx$ \zem. We simply summarize the main steps used to survey \CIV\ NALs in followings.

\begin{figure*}
\centering
\includegraphics[width=0.89\textwidth]{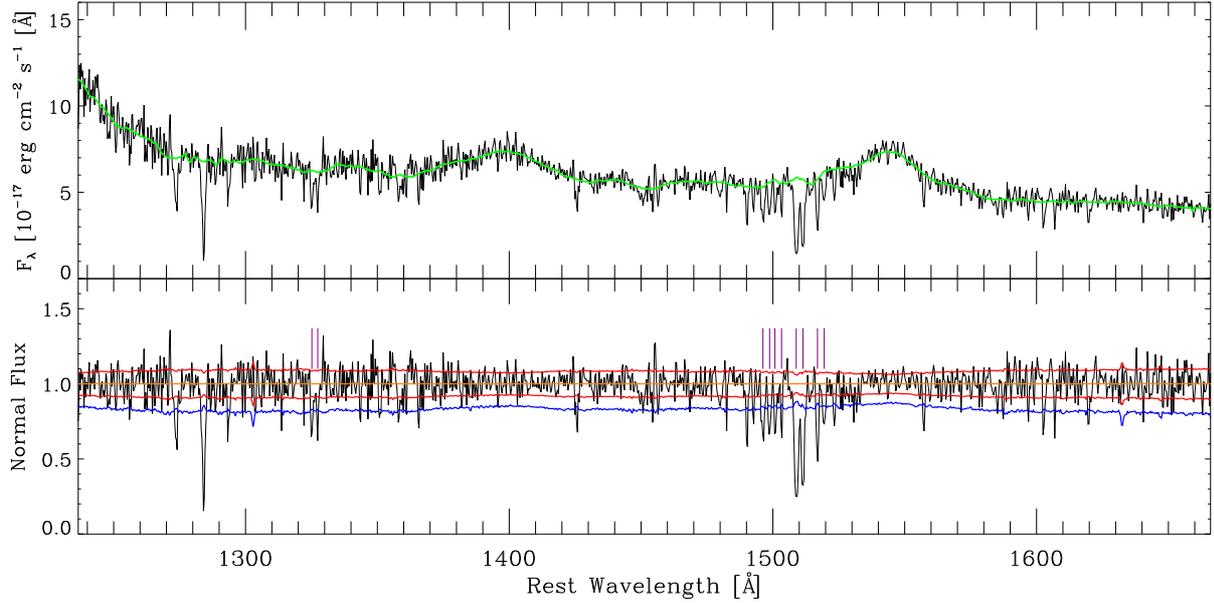}
\caption{Spectrum of quasar SDSS J081617.02+163640.3. Upper panel: the pseudo-continuum fitting (green) is overplotted on the quasar spectrum. Lower panel: the quasar spectrum divided by the pseudo-continuum fitting (normalized spectrum). The red and blue lines are the $\rm \pm1\sigma$ and $\rm -2\sigma$ flux uncertainty levels normalized by the pseudo-continuum fitting, respectively. The detected \CIV\ NALs with $S.L.\ge2$ and $W_r\ge0.2$ \AA\ are marked with purple lines.}
\label{fig2}
\end{figure*}

\begin{enumerate}
  \item Considering the underlying continuum and broad emission lines, a combination of cubic spline and Gaussian functions is accepted to fit a pseudo-continuum for each spectrum in iterative model \cite[e.g.,][]{2005ApJ...628..637N,2011AJ....141..137Q}. As an example, we overplot the pseudo-continuum fitting on the quasar spectrum with green line in Figure \ref{fig2}. The normalized spectrum is the quasar spectrum divided by the pseudo-continuum fitting, which is displayed with black line in the lower panel of Figure \ref{fig2}. The red and blue lines displayed in Figure \ref{fig2} are $\rm \pm1$\sig and -2\sig\ flux uncertainty levels divided by the pseudo-continuum fittings, respectively. We disregard absorption features above $\rm -2\sigma$ and continual absorption troughs (BALs) with line widths larger than 2000 \kms\ at depths greater than ten percent below the pseudo-continuum fitting.
  \item \CIV\ candidates are surveyed on the normalized spectrum. We invoke a pair of Gaussian functions to fit each candidate, and visually check one by one. NALs often show a line width of less than a few hundred \kms. Line widths of mini-BALs are between those of NALs and BALs. We conservatively reject the absorption candidates with $\rm FWHM \ge 800$ \kms, which are possible mini-BALs and beyond the scope of a series of works of \CIV\ NALs.
  \item Equivalent width at rest frame (\Wr) and corresponding uncertainty ($\sigma_{\rm w}$). The \Wr\ of each candidate is derived by integrating the Gaussian fitting. Based on the Gaussian fitting, the $\sigma_{\rm w}$ is evaluated via
      \begin{equation}
      \label{eq1}
      \sigma_w=\frac{\sqrt{\sum\limits_{i=1}^{N}P^2(\lambda_i-\lambda_0)\sigma^2_{f_i}\Delta\lambda_i^2}}{(1+z)\times\sum\limits_{i=1}^{N}P^2(\lambda_i-\lambda_0)},
      \end{equation}
      where $\Delta\lambda_i$, $P(\lambda_i-\lambda_0)$, $\sigma_{f_i}$, and $N$ are the wavelength interval spanned by the pixel, line profile centered at $\lambda_0$, flux uncertainty normalized by pseudo-continuum fitting, and the number of pixels over $\pm3\sigma$, here $\sigma$ is given by the Gaussian fitting of a certain candidate.
  \item Significant level of absorption line (here after $S.L.$). On the normalized spectrum, in terms of the largest depth ($S_{\rm abs}$) of the Gaussian fitting relative to unity and the mean value ($\sigma_{\rm S}$) of the normalized flux uncertainty around the absorption feature over $\pm3\sigma$, the $S.L.$ of each absorption line candidate is obtained by
      \begin{equation}
      \label{eq2}
       S.L.=\frac{1-S_{abs}}{\sigma_S},
      \end{equation}
  \item \CIV\ NALs are selected from the candidates with $S.L.\ge 2.0$ for both lines of each doublets. We get rid of absorption lines with \Wr $<0.2$ \AA, no matter what values of S.L. are, as we did in Papers I and II. A caveat here is that, the minimum equivalent width limit of absorption line is not a criterion to search for absorption candidates for our program. In other words, our sample completeness is determined by the $S.L.$ rather than the equivalent width (see Section \ref{sect3}).
\end{enumerate}

From 37,241 quasar spectra, we have detected a total number of 41,479 \CIV\ NALs with $1.4544\le$\zabs$\le4.9224$, of which 23,336 doublets with \zabs$\rm \ll$\zem\ are released in Papers I and II, and 18,143 doublets with \zabs$\rm \approx$\zem\ are newly detected in this paper. We distribute \zabs\ of these 41,479 \CIV\ NALs with back solid line in Figure \ref{fig1}, and list the properties of each absorption line in Table \ref{table1}. Here after, all analyses about \CIV\ NALs are referred to the whole \CIV\ NAL sample that includes doublets released in Papers I and II, and newly detected in this paper.

\begin{table*}
\centering
\caption{Catalog of \CIV narrow absorption line systems} \tabcolsep 1.1mm  \label{table1}
 \begin{tabular}{cccccccccccccc}
 \hline\hline\noalign{\smallskip}
SDSS NAME & PLATEID & MJD & FIBERID & $\rm z_{em}$ & $\rm z_{abs}$ & $\rm W_r^{\lambda1548}$ &$\rm \sigma_{W_r}^{\lambda1548}$& $\rm W_r^{\lambda1551}$ & $\rm \sigma_{W_r}^{\lambda1551}$ & $\rm S.L.^{\lambda1548}$ & $\rm S.L.^{\lambda1551}$ & $\rm \beta$\\
\noalign{\smallskip}
&&&&&&\AA&\AA&\AA&\AA&\\
\hline\noalign{\smallskip}
000001.93-001427.4 & 4216 & 55477 & 0312 & 2.1645  & 2.1625  & 0.31  & 0.08  & 0.24  & 0.05  & 3.87  & 4.19  & 0.00063  \\
000001.93-001427.4 & 4216 & 55477 & 0312 & 2.1645  & 2.1455  & 1.41  & 0.09  & 0.94  & 0.06  & 14.16  & 14.37  & 0.00602  \\
000003.17+011510.6 & 4296 & 55499 & 0364 & 2.3543  & 1.8863  & 0.29  & 0.10  & 0.36  & 0.14  & 2.74  & 2.50  & 0.14915  \\
000014.07+012951.5 & 4296 & 55499 & 0370 & 3.2284  & 2.6728  & 0.70  & 0.16  & 0.68  & 0.17  & 4.23  & 3.77  & 0.13994  \\
000015.17+004833.2 & 4216 & 55477 & 0718 & 3.0277  & 2.5222  & 1.11  & 0.16  & 0.84  & 0.11  & 6.39  & 7.10  & 0.13331  \\
000015.17+004833.2 & 4216 & 55477 & 0718 & 3.0277  & 2.9869  & 0.27  & 0.10  & 0.26  & 0.10  & 2.67  & 2.50  & 0.01018  \\
000017.97+002426.9 & 4216 & 55477 & 0724 & 2.2825  & 2.2648  & 1.83  & 0.14  & 1.39  & 0.12  & 12.68  & 10.63  & 0.00541  \\
000018.18+050803.6 & 4277 & 55506 & 0862 & 2.2278  & 2.1529  & 1.12  & 0.18  & 0.77  & 0.25  & 5.75  & 2.96  & 0.02347  \\
000018.64+033254.1 & 4277 & 55506 & 0106 & 2.2375  & 2.2323  & 0.46  & 0.10  & 0.43  & 0.15  & 4.47  & 2.79  & 0.00161  \\
000027.01+030715.5 & 4296 & 55499 & 0630 & 2.3533  & 1.9833  & 0.22  & 0.05  & 0.22  & 0.05  & 3.87  & 4.36  & 0.11639  \\
000027.01+030715.5 & 4296 & 55499 & 0630 & 2.3533  & 2.1303  & 0.91  & 0.04  & 0.69  & 0.04  & 20.34  & 18.25  & 0.06871  \\
\hline\hline\noalign{\smallskip}
\end{tabular}
\\
\begin{flushleft}
 Note: $\rm \beta=v/c=((1+z_{em})^2-(1+z_{abs})^2)/((1+z_{em})^2+(1+z_{abs})^2)$.
\end{flushleft}
\end{table*}

\section{Completeness and redshift path}
\label{sect3}
\begin{figure}
\centering
\includegraphics[width=7.5cm,height=5.8cm]{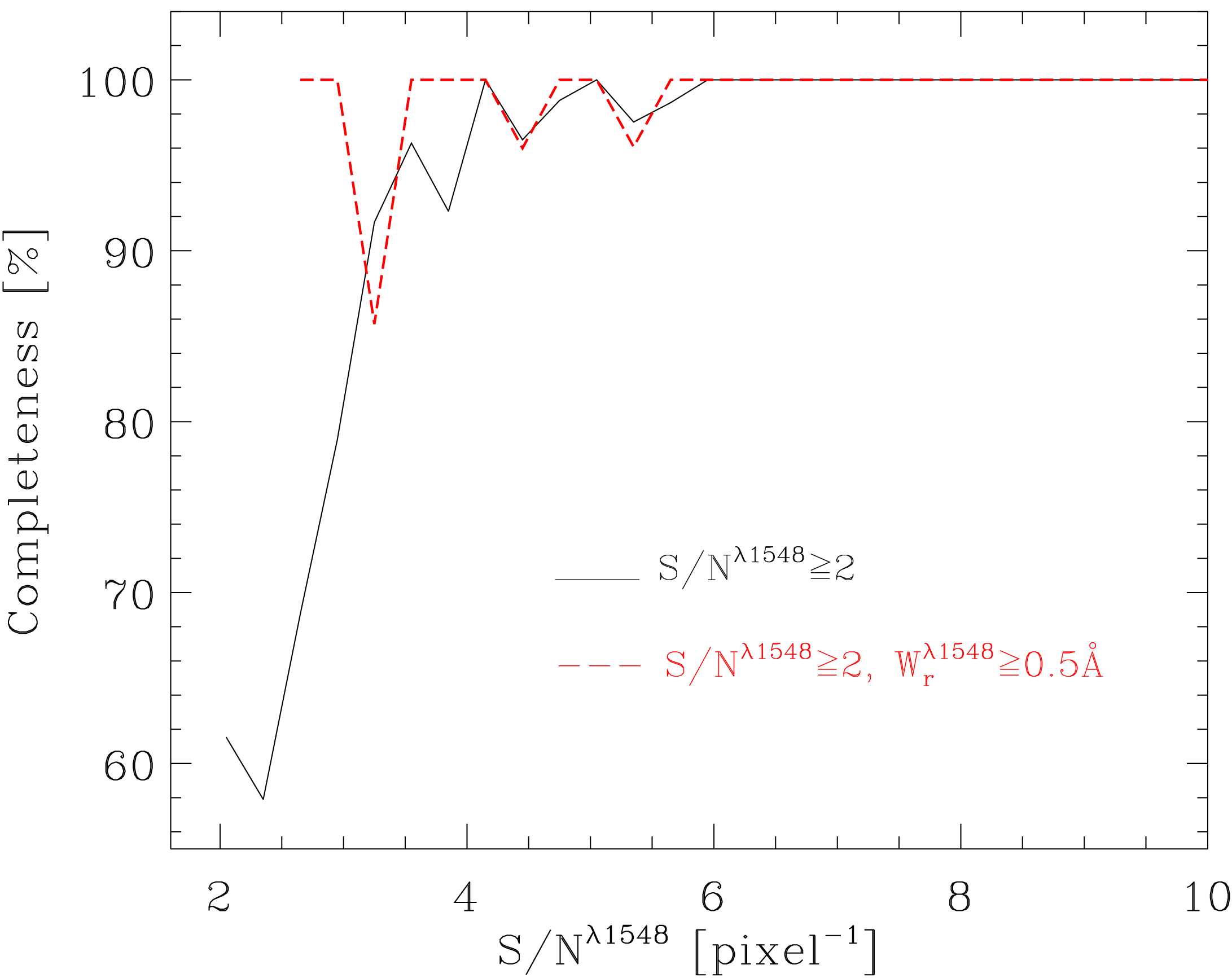}
\caption{Completeness as a function of $S/N^{\lambda1548}$. The black line accounts for the systems with $S/N^{\lambda1548}\ge2$, and the red dash line only considers the systems with $S/N^{\lambda1548} \ge2$ and \Wr$\ge0.5$\AA. The detection tends to be complete when $S/N^{\lambda1548}\ge4$.}
\label{fig3}
\end{figure}

We run Monte Carlo simulation to test the efficiency of detection rate of absorbers with different signal-to-noise ratio. Using the spectral data of the BOSS quasar, we simulate 1000 \CIV\ NALs with different $S.L.$. Then, based on the spectral resolution of the BOSS, we randomly insert simulated \CIV\ doublets onto 400 quasar spectra, where the detected \CIV\ systems have been marked out, in the surveyed spectral region of \CIV\ NALs. Using the method and criteria described in Section \ref{sect2}, we collect the simulated doublets from the 400 quasar spectra. At a given $S.L.$, we define the completeness of detection as the ratio of the number of detected doublets to that of simulated doublets. The result is plotted in Figure \ref{fig3}, which suggests that the detectability tends to be complete in range of $S/N^{\lambda1548}>4$.

\begin{figure*}
\centering
\includegraphics[width=7.5cm,height=5.8cm]{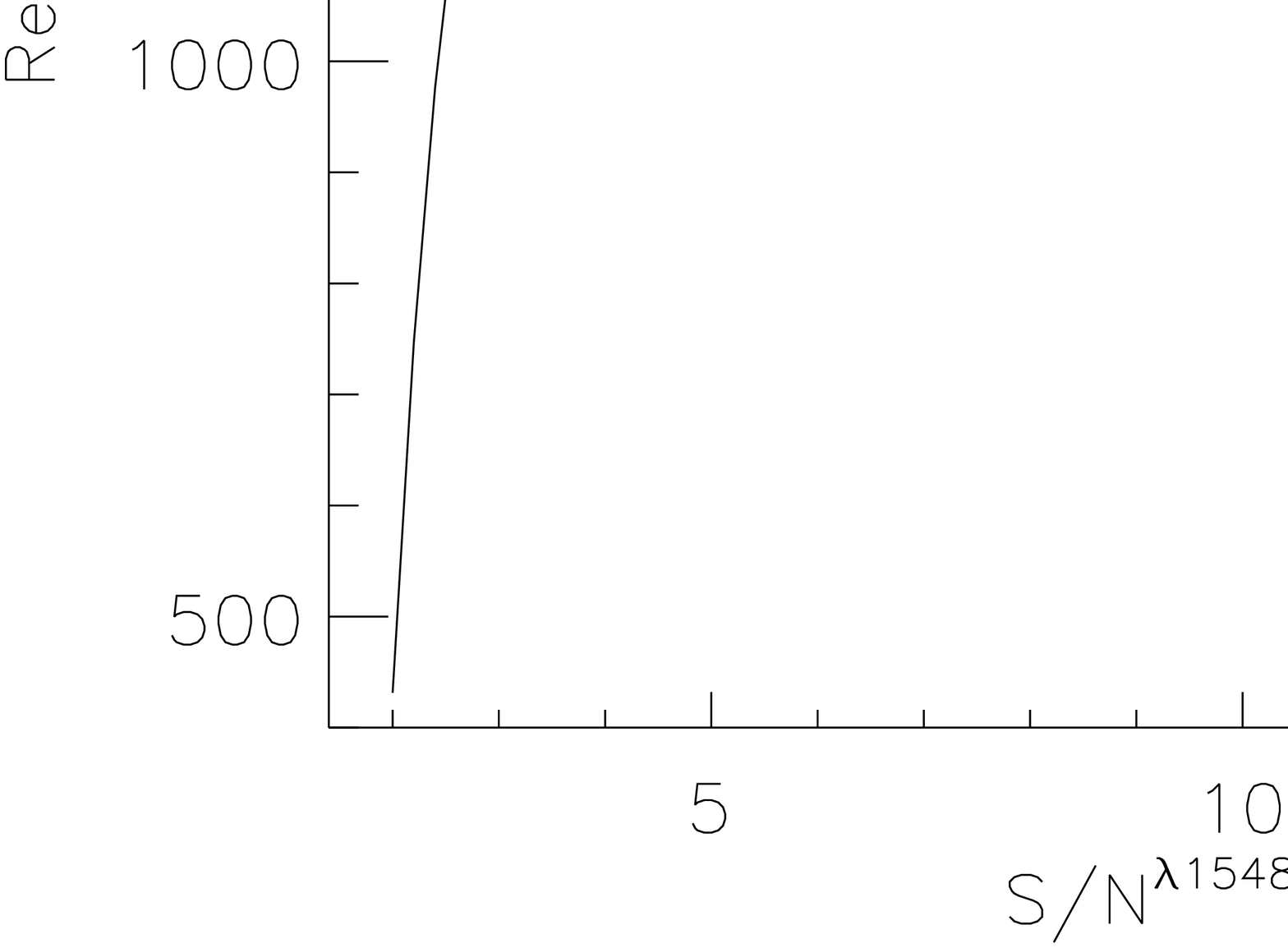}
\hspace{5ex}
\includegraphics[width=7.5cm,height=5.8cm]{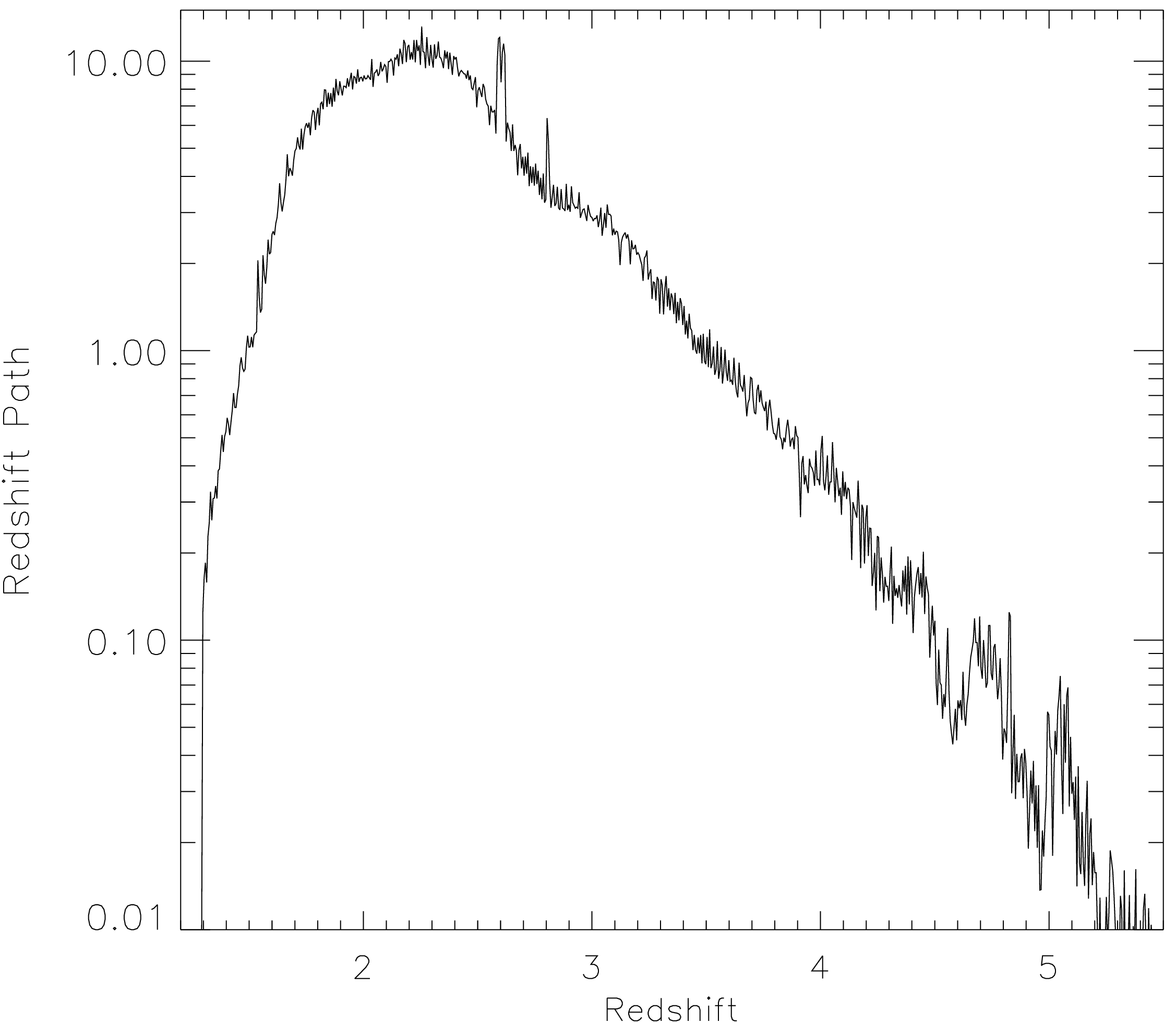}
\caption{Left panel: cumulative redshift path versus $S/N^{\rm \lambda1548}$. Right panel: total redshift path versus redshift. The pinnacles at $\rm z \approx 2.6$ and 2.8, and the obvious features at $\rm z > 4.2$ are likely due to poor subtractions of sky lines of \OI\ and $\rm Na~D$, and $\rm OH$ lines in the SDSS spectra.}
\label{fig4}
\end{figure*}

In surveyed spectral region, the total redshift path covered by surveys as a function of signal-to-noise ratio per pixel ($S/N^{\rm \lambda1548}$) under the pseudo-continuum fittings is derived by
\begin{equation}
\label{eq3}
\Delta Z(S/N^{\lambda1548})=\sum_{i=1}^{N_{spec}}\int_{z_i^{min}}^{z_i^{max}}g_i(S/N^{\lambda1548},z)dz,
\end{equation}
where, the sum is over all quasar spectra used to survey absorption lines, the integral boundaries ([$z_i^{\rm min}$,$z_i^{\rm max}$]) are constrained by surveyed spectral region, the sensitivity function $g_i(S/N^{\rm \lambda1548},z)=1$ if detection threshold $\rm S/N^{lim}~\le S/N^{\rm \lambda1548}$, otherwise $\rm g_i(S/N^{\rm \lambda1548},z)=0$. In the left panel of Figure \ref{fig4}, we display the cumulative redshift path as a function of $S/N^{\rm \lambda1548}$. The right panel of Figure \ref{fig4} shows the total redshift path as a function of redshift. The pinnacles at $\rm z \approx 2.6$ and 2.8, and the obvious features at $\rm z > 4.2$ are likely due to poor subtractions of sky lines, for example, \OI\ and $\rm Na~D$, and $\rm OH$ lines.

The spectral regions with strong emission lines often have higher signal-to-noise ratio than those without. Thus, for our identification procedure, the detection rate of \CIV\ doublet per redshift path could be higher for systems with \zabs$\rm \approx$\zem, which are located on or near to \CIV\ emission lines, than those with \zabs$\rm \ll$\zem. This should be noted in the following statistics and discussion.

\section{Equivalent width distribution and saturability}
\label{sect4}
\begin{figure}
\centering
\includegraphics[width=7.5cm,height=5.8cm]{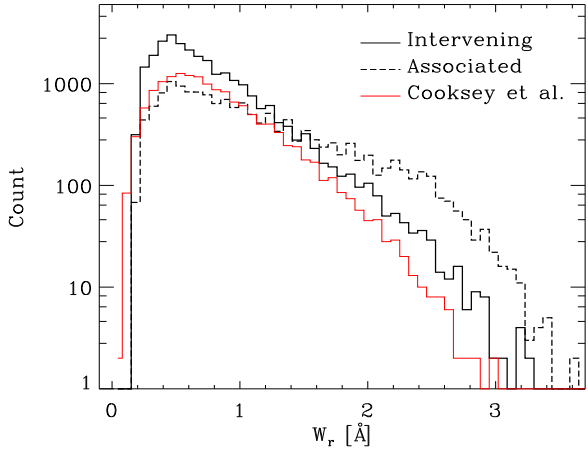}
\caption{\Wr\ distributions of \CIV\ NALs. The thick black solid line represents \CIV\ NALs with $\upsilon\equiv\beta c >5000$ \kms\ (intervening) included in our catalog, the dash line represents those with $\upsilon\equiv\beta c \le5000$ \kms\ (associated) included in our catalog, and the thin red solid line represents \CIV\ NALs of \citet{2013ApJ...763...37C}. A break near 0.5\AA is obvious.}
\label{fig5}
\end{figure}

The equivalent width of absorption line is an important parameter to reflect absorption strength and elemental abundance of absorber. In our \CIV\ NAL catalog, about 72\% of absorbers have $W_{\rm r}^{\rm \lambda1548}\ge0.5$\AA, $\sim$29\% have $W_{\rm r}^{\rm \lambda1548}\ge1$\AA, and $\sim$5\% have $W_{\rm r}^{\rm \lambda1548}\ge2$\AA. Using the SDSS quasar spectra, \cite{2013ApJ...763...37C} have assembled a catalog of intervening \CIV\ NALs with $\upsilon\equiv\beta c >5000$ \kms. In order to compare our detection to that of \cite{2013ApJ...763...37C}, we also use the criterion of $\upsilon\equiv\beta c =5000$ \kms\ to divide our \CIV\ NAL catalog into two subcatalogs of associated and intervening \CIV\ NALs. Figure \ref{fig5} shows the distributions of the \Wr, which is clear that the absorptions of associated systems tend to be stronger than those of intervening ones (the black solid and dash lines). Figure \ref{fig5} also indicates that the \Wr\ distribution of intervening systems included in our catalog is similar to that of \cite{2013ApJ...763...37C} (the red and black solid lines). These distributions show a break around $W_{\rm r}^{\rm \lambda1548}=0.5$\AA, which might suggest an uncomplete detection for systems with $W_{\rm r}^{\rm \lambda1548}<0.5$\AA. The weak systems often have small $S.L.$, and the uncomplete detection suggested by the black solid line of Figure \ref{fig3} might be mainly introduced by systems with $W_{\rm r}^{\rm \lambda1548}<0.5$\AA. When the Monte Carlo simulation for estimating the detected completeness (see Section \ref{sect3}) is limited to systems with $W_{\rm r}^{\rm \lambda1548}\ge0.5$\AA, we show the detected completeness as a function of $S/N^{\rm \lambda1548}$ with red dash line in Figure \ref{fig3}, which implies that the limit of $W_{\rm r}^{\rm \lambda1548}\ge0.5$\AA\ does not change the cut, namely $S/N^{\rm \lambda1548}>4$, of the detected completeness.

The saturability of resonant absorption doublet can be obtained by the ratio of absorption strength \cite[$\rm DR$,][]{1948ApJ...108..242S}. For \CIV\ resonant doublet, on the theoretical hand, $\rm DR\equiv W_r^{\rm\lambda1548}/W_r^{\rm \lambda1551} $ can vary from $1$ for full saturation to $2$ for full unsaturation. The error of the $\rm DR$ can be evaluated via
\begin{equation}
\label{eq4}
\sigma_{\rm DR} \equiv \Delta {\rm DR} = \sqrt{ (\frac{\Delta W_r^{\rm \lambda1548}}{W_r^{\rm \lambda1551}})^2 + (\frac{W_r^{\rm \lambda1548}\Delta W_r^{\rm \lambda1551}}{(W_r^{\rm \lambda1551})^2})^2 }
\end{equation}

\begin{figure}
\centering
\includegraphics[width=7.5cm,height=5.8cm]{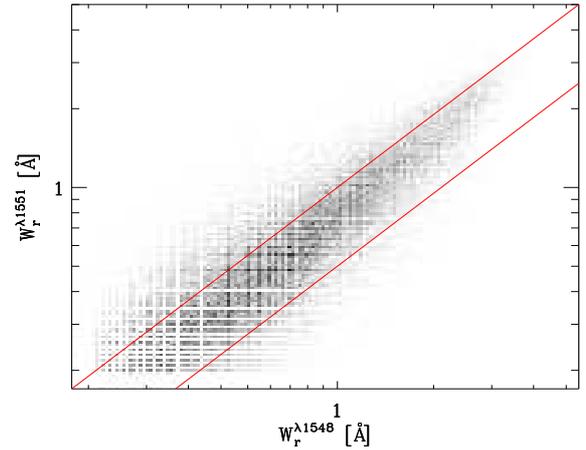}
\caption{Local data point densities with grayscale. The red lines represent ${\rm DR} = W_r^{\rm\lambda1548}/W_r^{\rm \lambda1551} = 1$ and 2, which correspond the theoretical limits of full saturation and unsaturation, respectively.}
\label{fig6}
\end{figure}

Figure \ref{fig6} displays $W_{\rm r}^{\rm\lambda1551}$ vs $W_{\rm r}^{\rm\lambda1548}$. We find that about 73\% of \CIV\ NALs fall into the range of $1\le {\rm DR}\le2$, and only $\sim$ 7.7\% of \CIV\ NALs are beyond the range of $1-\sigma_{\rm DR}\le {\rm DR}\le2+\sigma_{\rm DR}$. Meanwhile, at $W_{\rm r}^{\rm \lambda1548} >1$\AA, the $\rm DR$ monotonically declines with increasing $W_{\rm r}^{\rm\lambda1548}$, indicating that the stronger absorptions correspond to the more saturated absorptions that follows the curve of growth.

\section{Discussion}
\label{sect5}
\subsection{The redshift density evolution of absorbers}
\begin{figure}
\centering
\includegraphics[width=7.cm,height=6.cm]{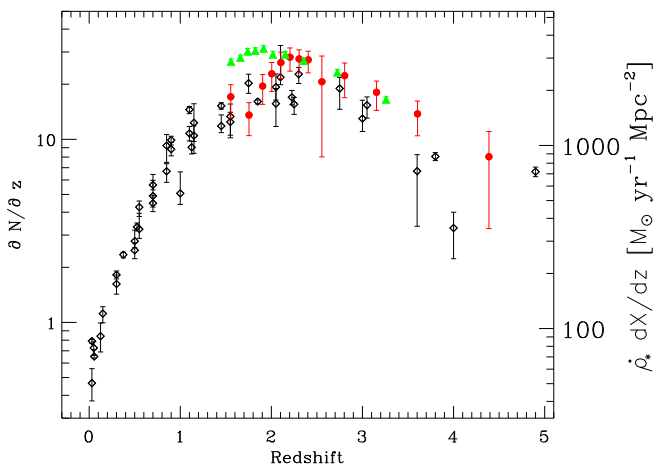}
\caption{The reshift density evolution of absorbers from two surveys: this work (\CIV\ NALs, red filled circles), \citet{2013ApJ...763...37C}(\CIV\ NALs, green filled triangles). The star formation rate densities ($\dot{\rho}_\ast$) based on the data from \citet{2014ARA&A..52..415M} are displayed with black unfilled diamonds. The mappings between the left-axes of the triangles and the circles, and between the left and right y-axes are arbitrary.}
\label{fig7}
\end{figure}

Systematical survey of NALs on spectra of large quasar sample is in favor of studies of the absorber evolution and the connection between absorbers and galaxies. The redshift density is the count of absorbers with specified limits and normalized by redshift path length. In a specified redshift range, the redshift density of absorbers with a specified interval of $S/N^{\rm \lambda1548}$ is derived via
\begin{equation}
\label{eq5}
\frac{\partial N}{\partial z}=\sum_i^{N_{\rm abs}}\frac{1}{\Delta Z(S/N^{\rm \lambda1548}_i)}
\end{equation}
where $\Delta Z(S/N^{\rm \lambda1548})$ is calculated by Equation \ref{eq3}, and the sum is over all \CIV\ NALs. Here, we only propagate the uncertainty of $\Delta Z(S/N^{\rm \lambda1548})$ into the uncertainty of $\partial N/\partial z$. Using the spectral flux uncertainty limited by the SDSS, we invoke the Monte Carlo simulation to resample the spectral flux, and then limit the uncertainty of $\Delta Z(S/N^{\rm \lambda1548})$ from Poisson statistics. Thus, the uncertainty of $\partial N/\partial z$ can be simply propagated by
\begin{equation}
\label{eq6}
\Delta (\partial N/\partial z) =\frac{N_{\rm abs}\times\Delta (\Delta Z(S/N^{\rm \lambda1548}))}{\Delta Z(S/N^{\rm \lambda1548})^2}
\end{equation}
In several redshift ranges and at $S/N^{\rm \lambda1548}\ge 2$, Table \ref{table2} lists the counts of \CIV\ NALs, $\Delta Z(S/N^{\rm \lambda1548})$, and $\partial N/\partial z$, and Figure \ref{fig7} displays the distribution of $\partial N/\partial z$ as a function of redshift with red filled circles. We also show the $\partial N/\partial z$ of \cite{2013ApJ...763...37C} with green filled triangles in Figure \ref{fig7}, which are measured from quasar spectra of the SDSS-I/II. It is clear that the $\partial N/\partial z$ of ours and \cite{2013ApJ...763...37C} have parallel evolution profile. The specified values of $\partial N/\partial z$ depend on methodologies used to determine redshift path. Thus $\partial N/\partial z$ related to $\Delta Z(S.L.)$ cannot be compared directly to $\partial N/\partial z$ related to $\Delta Z(W_{\rm r}^{\rm \lambda})$. The redshift path of this work is limited by $S.L.$, while that of \cite{2013ApJ...763...37C} is constrained by $W_{\rm r}^{\rm \lambda}$. Therefore, for comparison, the $\partial N/\partial z$ of \cite{2013ApJ...763...37C} plotted in Figure \ref{fig7} has been scaled arbitrarily.

\begin{table}
\caption{The redshift number density with $S/N^{\rm \lambda1548}\ge 2$} \tabcolsep 3mm \centering \label{table2}
 \begin{tabular}{cccccccccccccc}
 \hline\hline\noalign{\smallskip}
z bin & $N_{\rm abs}$ & $\rm \Delta Z$ & $\rm \partial N/\partial z$\\
\hline\noalign{\smallskip}
$[1.4544,1.6544)$ &1123 &65.8 $^{+15.5}_{-10.8}$ &17.069$^{+2.804}_{-4.009}$\\
$[1.6544,1.8544)$ &2797 &206.2$^{+46.9}_{-34.9}$ &13.569$^{+2.299}_{-3.092}$\\
$[1.8544,1.9544)$ &2786 &142.7$^{+29.3}_{-22.3}$ &19.530$^{+3.058}_{-4.005}$\\
$[1.9544,2.0544)$ &3604 &158.1$^{+31.4}_{-24.1}$ &22.801$^{+3.479}_{-4.532}$\\
$[2.0544,2.1544)$ &4423 &168.5$^{+30.7}_{-23.6}$ &26.249$^{+3.683}_{-4.783}$\\
$[2.1544,2.2544)$ &5444 &193.8$^{+31.2}_{-23.9}$ &28.093$^{+3.465}_{-4.519}$\\
$[2.2544,2.3544)$ &4902 &178.4$^{+28.5}_{-21.6}$ &27.484$^{+3.330}_{-4.392}$\\
$[2.3544,2.4544)$ &4142 &152.4$^{+23.1}_{-17.4}$ &27.176$^{+3.113}_{-4.111}$\\
$[2.4544,2.6544)$ &4897 &237.4$^{+145.1}_{-90.6}$ &20.634$^{+7.880}_{-12.614}$\\
$[2.6544,2.9544)$ &3817 &171.5$^{+41.5}_{-29.7}$ &22.256$^{+3.861}_{-5.381}$\\
$[2.9544,3.3544)$ &2465 &136.4$^{+27.9}_{-20.6}$ &18.070$^{+2.735}_{-3.691}$\\
$[3.3544,3.8544)$ &866  &62.9 $^{+15.2}_{-11.0}$ &13.780$^{+2.415}_{-3.325}$\\
$[3.8544,4.9224]$ &212  &26.4 $^{+15.7}_{-9.8} $ &8.032$^{+2.996}_{-4.778}$\\
\hline\hline\noalign{\smallskip}
\end{tabular}
\begin{flushleft}
Note --- $N_{\rm abs}$ represents the number of detected \CIV\ NALs in corresponding redshift intervals.
\end{flushleft}
\end{table}

The connection between the metal absorptions and the star formation rate densities has been widely explored in many previous works \cite[e.g.,][]{2011MNRAS.418.2730G,2011MNRAS.417..801M,2013ApJ...773...16Z,2016MNRAS.455.1713H}. The cosmic star formation history peaks at $z\approx2$ and decreases by about one magnitude from $z\sim1$ to $z\sim0$ \cite[see recent review][]{2014ARA&A..52..415M}. In order to establish the connection between  $\partial N/\partial z$ and cosmic star formation histories, we scale the star formation rate densities ($\dot{\rho}=dM_\ast/(dtdAdX)$) with $dX/dz$, where X is the comoving distance and A is the comoving area. The scaling $\dot{\rho}$ is plotted with open diamonds in Figure \ref{fig7}. The global evolution shapes of $\dot{\rho}dX/dz$ and $\partial N/\partial z$ are very similar.

The redshift number density $\partial N/\partial z$\footnote{More than 60\% C IV NALs show $DR<1+\sigma_{\rm DR}$, which means that most of \CIV\ NALs tend to be saturated absorptions. The column density derived from the curve of growth would intrinsically show large uncertainty for saturated NALs, which leads to large uncertainties in the estimation of mass density of \CIV\ absorbing gas.} relates to the volume density of absorbers $n_{\rm vol}$ and their physical cross-section $\sigma_{\rm phys}$. Therefore, the evolution of $\partial N/\partial z$ is possibly dominated by the product of $n_{\rm vol}\sigma_{\rm phys}$. \CIV\ metal absorbing gas relates to galaxy \cite[e.g.,][]{2001ApJ...556..158C,2005ApJ...629..636A,2008ApJ...679.1218T,2010ApJ...721..174M,2013ApJ...779L..17B,2014ApJ...796..136B}. Both the number of galaxies and their surrounding gas (halo) likely evolve over cosmic time, so evolutions in both $n_{\rm vol}$ and $\sigma_{\rm phys}$ can change the value of $\partial N/\partial z$:
\begin{equation}
\label{eq:cross_section}
\frac{\partial N}{\partial z} \propto n_{\rm vol}(z)\sigma_{\rm phys}(z).
\end{equation}
Using the UV luminosity functions from \cite{2007ApJ...670..928B}, \cite{2009ApJ...692..778R} and \cite{2010ApJ...725L.150O}, \cite{2013ApJ...763...37C} derived the UV-selected galaxy volume density ($n_{\rm vol}$) by integrating the best fitting Schechter function, which increase by a factor of 2 to 3 in redshift range of 5 to 1.5 (see Figure 12 of \cite{2013ApJ...763...37C}). In similar redshift range, Figure \ref{fig7} and Table \ref{table2} show that the $\partial N/\partial z$ increases by a factor of 3 to 4. In this case, the physical cross-section $\sigma_{\rm phys}$ of absorbing cloud could modestly increase over cosmic time. This increase might be from a comprehensive evolution of absorber's UV ionizing background, metallicity, and geometric size.

\subsection{The quasar-frame velocity distribution}
NALs are mainly arisen from: (1) material of intervening galaxies lying on quasar sightlines; and (2) material associated with quasar themselves, which can be related to quasars' host galaxies, winds/outflows and surrounding environment. The former produces intervening NALs with \zabs$\ll$\zem, and the latter brings about associated NALs with \zabs$\approx$\zem. Associated NALs are alluring since they provide a powerful tool to explore AGN feedback. BALs are unambiguously associated with quasars. However, it is difficult to identify whether one NAL is related to quasar or not. The line variability and partial coverage analysis are frequently applied to diagnose the associated NALs. Only small sample of variable NALs are detected up to nowadays \cite[e.g.,][]{2004ApJ...601..715N,2004ApJ...613..129W,2013MNRAS.434..163H,2015MNRAS.450.3904C}. Therefore, the statistics is often utilized to decompose associated population from a large NAL catalog \cite[e.g.,][]{2007ApJS..171....1M,2008MNRAS.386.2055N,2008MNRAS.388..227W}.

If NALs are associated to quasars, then the differences between \zabs\ and \zem\ should be dominated by motions of absorbers relative to quasars, and the relative velocities can be derived by
\begin{equation}
\label{eq7}
\beta\equiv\frac{\upsilon}{c}=\frac{(1+z_{\rm em})^2-(1+z_{\rm abs})^2}{(1+z_{\rm em})^2+(1+z_{\rm abs})^2}.
\end{equation}
We note that \zabs\ determined from NALs have high accuracy, while \zem\ measured from broad emission lines can show an uncertainty from 100 \kms\ to 3000 \kms\ \citep[e.g.,][]{2010MNRAS.405.2302H,2011ApJS..194...45S,2016ApJ...817...55S}. Using the near-infrared quasar spectra, \cite{2016ApJ...817...55S} revealed that, after accounting for measurement errors, redshifts estimated from broad \MgIIwave\ emission lines have an intrinsical uncertainty of about 200 \kms\ with respect to the redshifts determined by \OIIIb\ emission lines. This small uncertainty implies that \MgII\ emission line is a better indicator to estimate quasar systemic redshift relative to \CIII, \CIV, and \lya\ emission lines \citep[e.g.,][]{2010MNRAS.405.2302H,2011ApJS..194...45S}.

\begin{figure*}
\centering
\includegraphics[width=0.31\textwidth]{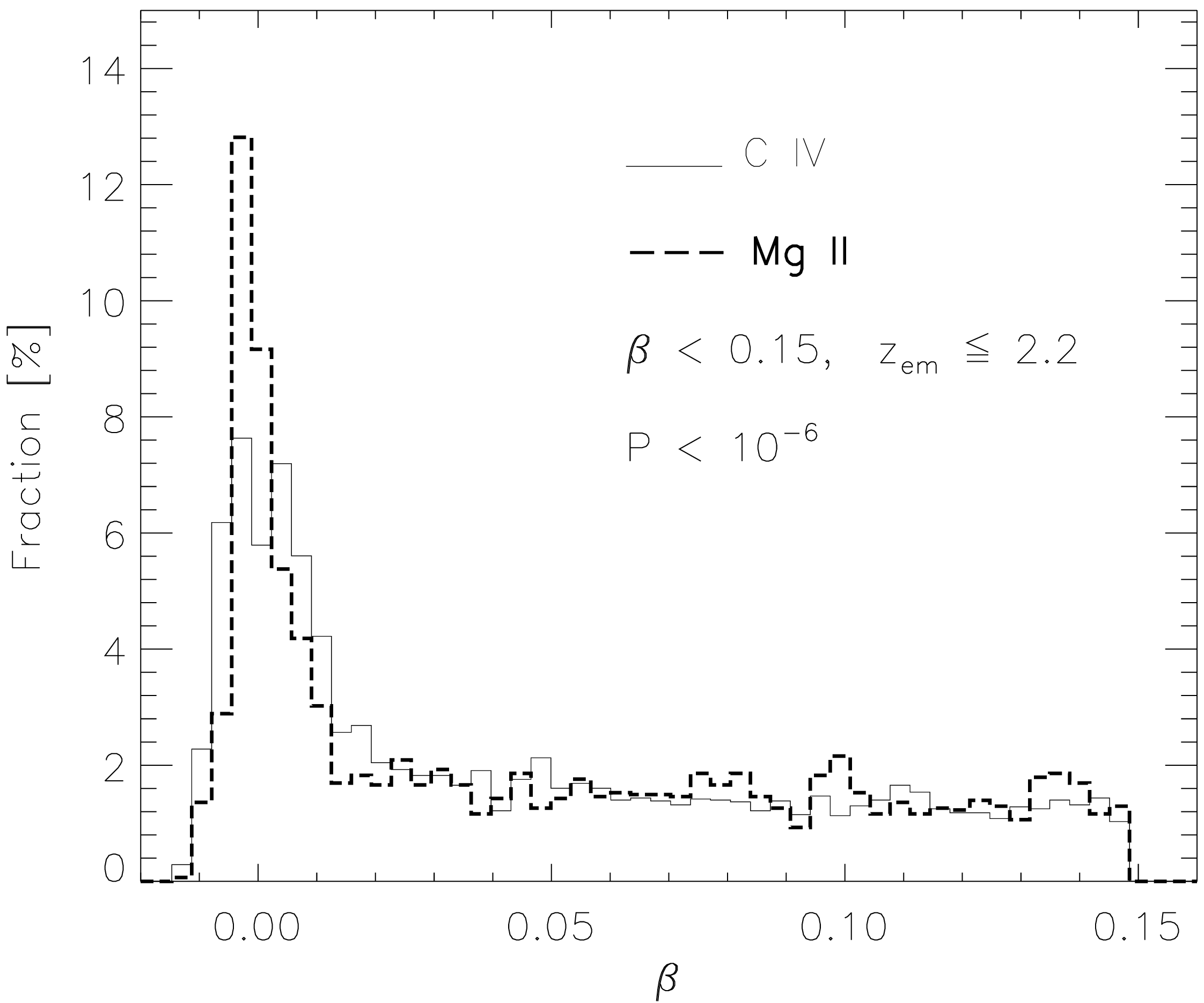}
\hspace{2ex}
\includegraphics[width=0.31\textwidth]{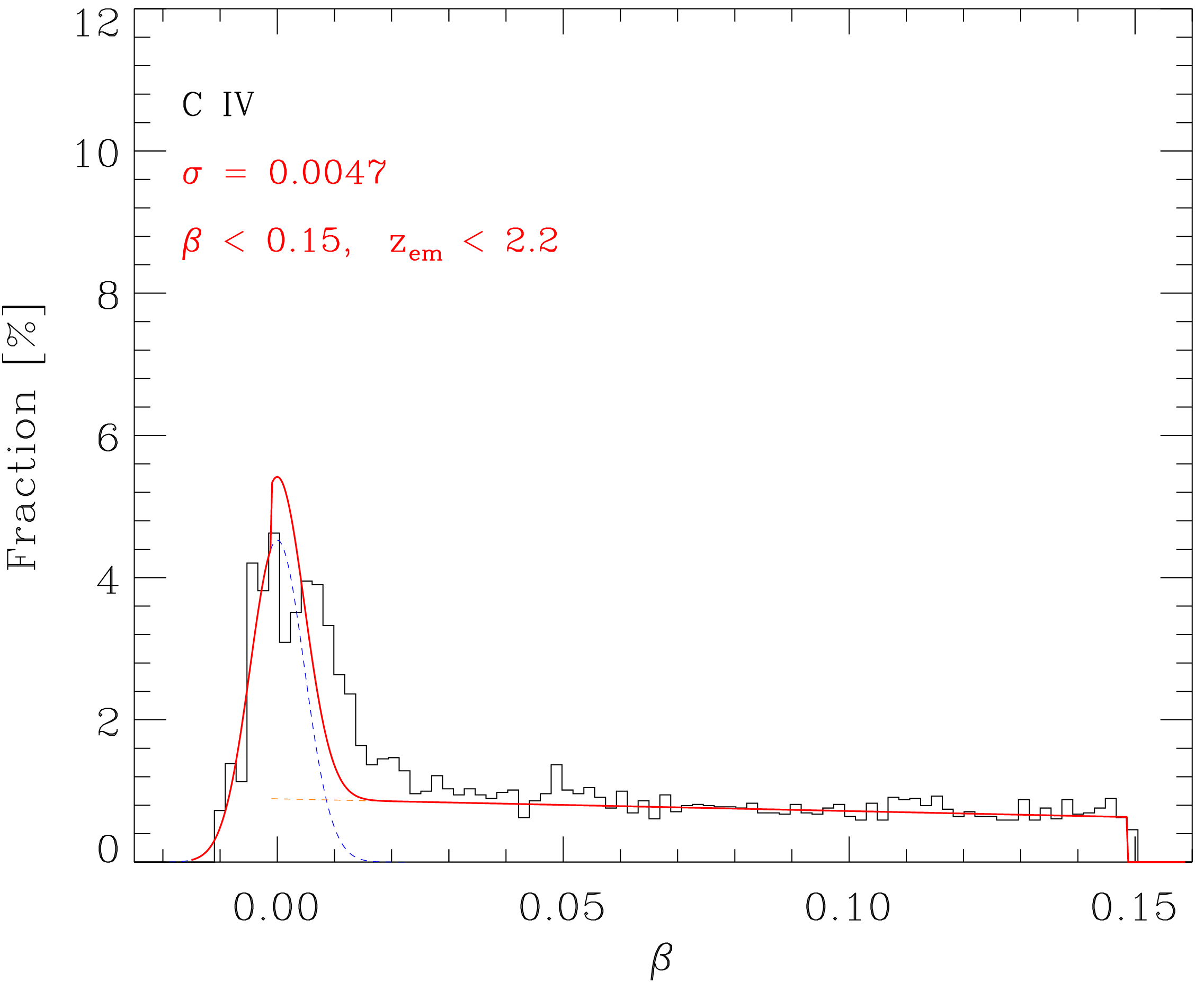}
\hspace{2ex}
\includegraphics[width=0.31\textwidth]{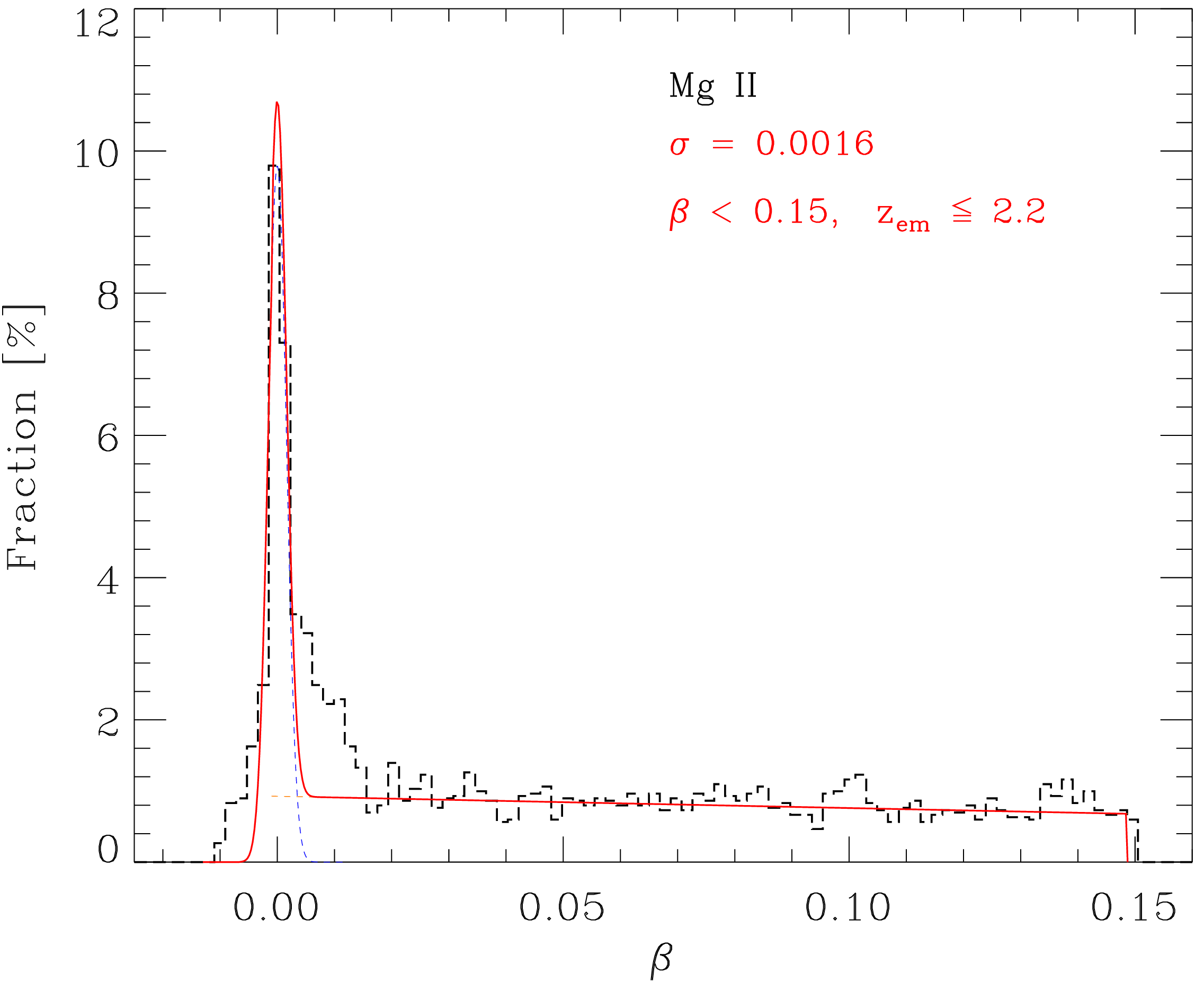}
\caption{Relative velocity distributions of absorbers with $\beta < 0.15$ and $z_{\rm em} \le 2.2$. The y-axes are the counts of absorbers normalized by the total numbers of absorbers contained in corresponding subsamples. The red curves are the reduced $\rm \chi^2$ fittings of a combination of a Gaussian function centred at $\beta = 0$ and a linear function using the data with $\beta<0$ and $>0.06$. The blue and orange dash lines are the Gaussian and linear components, respectively. Left panel: the KS test suggests a probability of $P<10^{\rm -6}$ that the two populations draw from the same parent sample. Middle panel: \CIV\ absorbers with $\sigma=0.0047$ for the Gaussian component. Right panel: \MgII\ absorbers with $\sigma=0.0016$ for the Gaussian component.}
\label{fig8}
\end{figure*}

\cite{2015ApJS..221...32C} has systematically surveyed \MgII\ NALs on 43,260 BOSS quasar spectra of the DR9Q from redward of \lya\ to red wing of \MgIIwave\ emission lines, which results in a total number of 18,598 \MgII\ doublets with $0.2933\le$\zabs$\le2.6529$. The method used in \cite{2015ApJS..221...32C} to search for NALs is exactly the same as this work. In order to compare the $\beta$ distribution of \CIV\ NALs to that of \MgII\ NALs, we make use of the \MgII\ NALs included in \cite{2015ApJS..221...32C}. We limit \CIV\ NALs to quasars with \zem$\le2.2$ so that \MgIIwave\ emission lines can be available by the BOSS spectra. Surveyed spectral region of \CIV\ NALs is between \lya\ and \CIVwave\ emission lines, which constrains $\beta<0.16$ for \CIV\ NALs. And that of \MgII\ NALs is between \lya\ and \MgIIwave\ emission lines, which permits measurement of much higher $\beta$ values for \MgII\ NALs. In order to make a consistent comparison, we limit both \CIV\ and \MgII\ NALs with $\beta<0.15$. Taking into account of \zem$\le2.2$ and $\beta<0.15$, there are 5922 \CIV\ NALs and  3012 \MgII\ NALs. We display the $\beta$ distributions of both \CIV\ and \MgII\ NALs in Figure \ref{fig8}.

In the calculation of $\beta$, we have made an assumption that all NALs are produced by quasar self-absorptions. In fact, the catalogs contain absorptions from (1) the first population: cosmologically intervening galaxies; (2) the second population: quasar host galaxies or galaxy clusters/groups that quasars reside in, which is often called as environmental absorptions; and (3) the third population: quasar outflows. Although ultra-fast outflows have been detected by many researchers \cite[e.g.,][]{2011MNRAS.410.1957H,2011ApJ...742...44T,2013ApJ...777...56C,2013ApJ...772...66G}, the vast majority of absorbers with a large $\beta$ belong to the first population. When absorbers fall into the second population, some absorbers move toward the observer  and some away, and the $\beta$ would show a normal distribution centred at $\beta=0$. In addition, galaxies around quasars are clustered/grouped, so one expects an excess of associated absorbers with respect to cosmologically intervening absorbers. Absorbers of the third population move towards the observer, thus they have $\beta>0$ and are expected to destroy the normal distribution of $\beta$ of the second population absorptions. Figure \ref{fig8} clearly shows that there are an abnormal excess in range of $\beta<0.02$ and a flat tail in range of $\beta>0.02$. This suggests that the data in range of $\beta>0.02$ would be dominated by absorptions of the first population, and those in range of $\beta<0.02$ be mainly contributed from absorptions of the second and third populations.

In theory, quasar outflows do not affect the $\beta$ distribution of associated absorbers with $\beta < 0$. In Figure \ref{fig8}, absorbers with $\beta > 0.06$ are probably contributed from cosmologically intervening absorptions. We explore a combination of a Gaussian function accounting for environmental absorptions and a linear function accounting for intervening absorptions to fit the data with $\beta<0$ and $>0.06$. This results in Gaussian components with $\sigma = 0.0047$ for \CIV\ NALs and $\sigma = 0.0016$ for \MgII\ NALs. The fitting results are plotted with color lines in Figure \ref{fig8}. It is clear that a combination of a Gaussian and a linear functions can not well characterize the overall distributions, and the excesses located at $\beta\approx0.01$ are very remarkable with respect to fitting curves, which likely are the contribution of quasar outflows.

Since \CIV\ (64.49 eV) has much higher ionization potential than \MgII\ (15.035 eV), we expect the associated \CIV\ and \MgII\ NALs to show different statistical properties. Recalling Figure \ref{fig8}, in range of $\beta<0.02$, the $\beta$ distribution of \MgII\ absorbers is more compact than the same values of \CIV\ absorbers. KS test derives a probability of $P<10^{\rm -6}$ that the $\beta$ of \CIV\ and \MgII\ absorbers draw from the same parent sample. The multi-component fittings overplotted with red lines imply that the dispersion of \CIV\ environmental absorptions ($\sigma=0.0047$) is much larger than the same value of \MgII\ environmental absorptions ($\sigma=0.0016$). In addition, with respect to multi-component fittings, the excessive tail of \MgII\ absorbers reaches a value of $\beta\approx0.02$, but that of \CIV\ absorbers can extend out to a higher value of $\beta>0.04$. This suggests that the higher ionization outflows have the larger maximum velocity.

Using \CIV\ and \MgII\ NALs detected from quasar spectra of the SDSS DR3, the model of \cite{2008MNRAS.388..227W} predicts that the scales that the quasar photoevaporates environmental absorbers are related to ionization energies of ions. The radius of 300 kpc for survived \CIV\ absorbers is much smaller than that of 800 kpc for survived \MgII\ absorbers. This discrepancy is also clearly reflected by our multi-component fittings (see Figure \ref{fig8}): $\sigma=0.0047$ for \CIV\ absorbers, and $\sigma=0.0016$ for \MgII\ ones.

Using the emission redshifts of the SDSS pipeline, \cite{2008MNRAS.386.2055N} obtained a dispersion with a value of 913 \kms\ for \CIV\ environmental absorbers after deconvolving the redshift uncertainty. When based on redshifts determined by \MgII\ emission lines, they found a value of about 450 \kms. For our multi-component fitting, we obtain a dispersion with a value of 728 \kms, which is slightly larger than that of 450 \kms. Our limit of \zem$\le2.2$ can reduce, on a certain extent, the uncertainties of emission redshifts. However, we can not guarantee that the significant \MgII\ emission line was imprinted on each quasar spectrum. Thus, the uncertainties of emission redshifts could be larger than those determined by \MgII\ emission lines.

\subsection{Dependence on the radio detection}
The observational area of the Faint Images of the Radio Sky at Twenty centimeter survey \cite[FIRST,][]{1995ApJ...450..559B} coincides with that of the SDSS. Over most of the area, the FIRST survey is limited to a detection threshold of 1 $\rm mJy$. In order to investigate the dependence of associated absorber properties on radio detection, we match our parent quasar sample with the FIRST within a radius of $30''$. Figure \ref{fig8} shows that some outflow absorbers can reach a high velocity. Here, we loosely constrain associated absorbers with $\beta<0.03$. Combining the limits of $\beta<0.03$ and \zem$\le2.2$, there are 2741 and 198 \CIV\ NALs imprinted on the spectra of FIRST undetected and detected quasars, respectively. In terms of the luminosity at 20 $\rm cm$, the quasar with a luminosity $>10^{\rm 25}~ \rm W/Hz$ is defined as radio-loud one, inversely radio-quiet one. According to this definition, the FIRST detected quasars of the 198 \CIV\ absorbers belong to radio-loud population. Because the small sample of the FIRST detected quasars, we do not discuss the dependence of associated absorber properties on radio morphologies.

\begin{figure}
\centering
\includegraphics[width=7.5cm,height=6.cm]{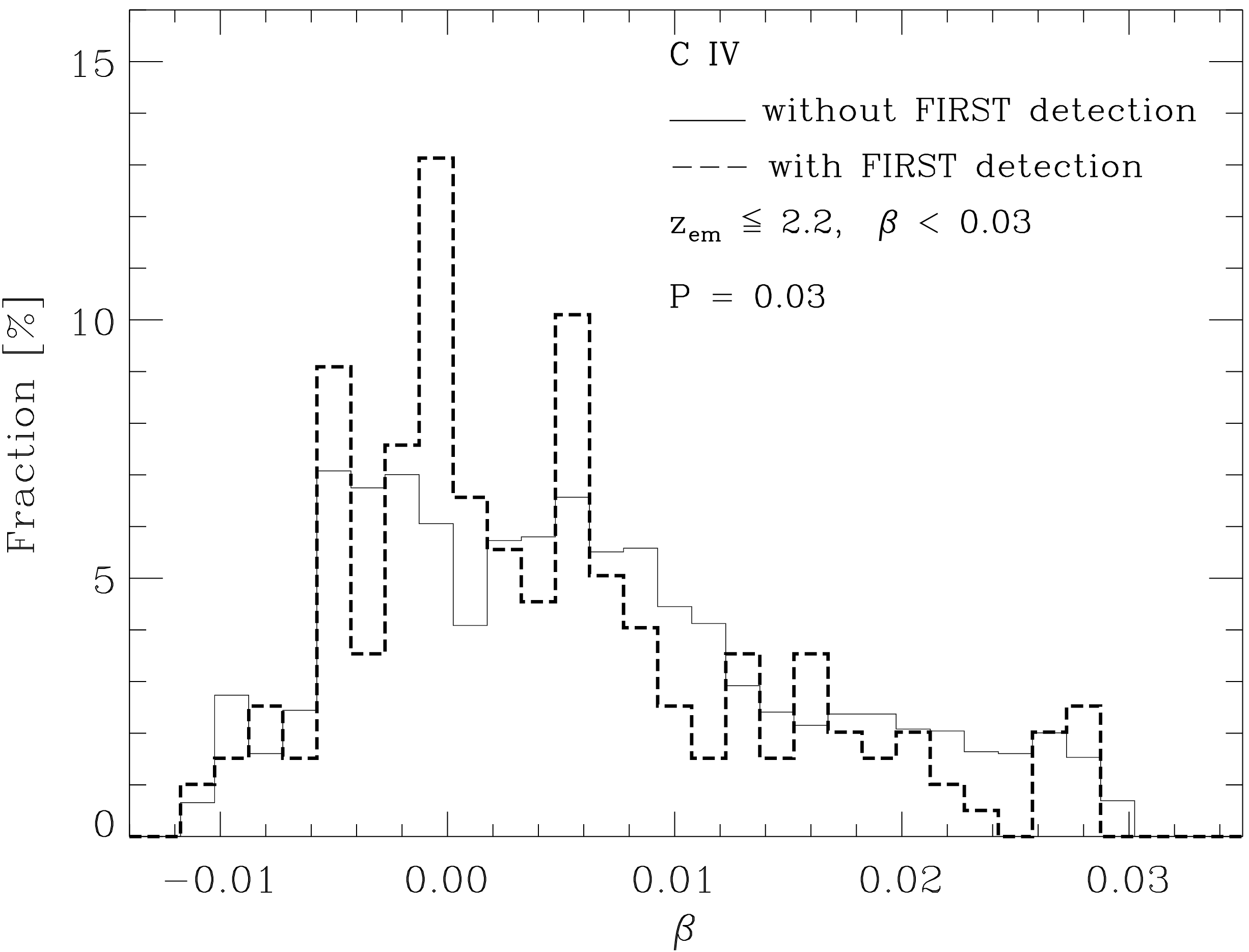}
\caption{Relative velocity distributions of absorbers with $\beta<0.03$ and \zem$\le2.2$.  The y-axis are the counts of absorbers normalized by the total numbers of absorbers contained in corresponding subsamples. The solid (dash) lines represent the quasars without (with) the FIRST detection. KS test yields a $P=0.03$}
\label{fig9}
\end{figure}

\begin{figure}
\centering
\includegraphics[width=7.5cm,height=6.cm]{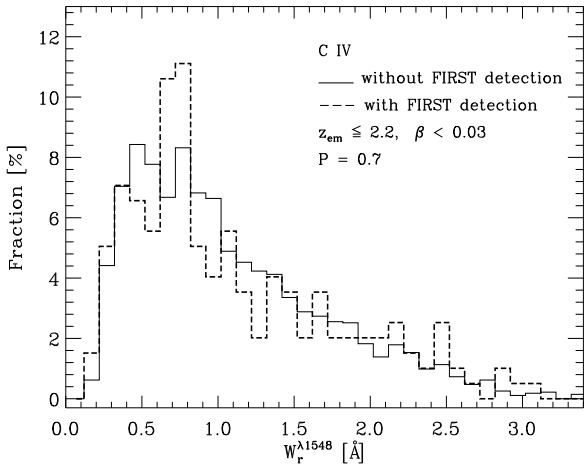}
\caption{\Wr\ distributions of absorbers with $\beta<0.03$ and \zem$\le2.2$. The y-axis are the counts of absorbers normalized by the total numbers of absorbers contained in corresponding subsamples. The solid (dash) lines represent the quasars without (with) the FIRST detection. KS test yields a $P=0.7$}
\label{fig10}
\end{figure}

We display $\beta$ distributions in Figure \ref{fig9}. No significant difference is shown between the $\beta$ of the FIRST detected and undetected NALs. This similar result is also exhibited in the \Wr\ distributions (see Figure \ref{fig10}). The $\beta$ distributions do not coincide with those of \cite{2008MNRAS.388..227W}, who found that the $\beta$ measured from radio-loud quasar spectra are more compact at $\beta\approx0$ than the same values measured from radio-quiet ones. There are only 69 FIRST detected quasars satisfied our criteria. This number is too small to allow us to further investigate the different $\beta$ distributions between our results and \cite{2008MNRAS.388..227W}.

\subsection{Dependence on the UV continuum luminosity}
The ionizing radiation from quasars produces feedback to surrounding gas, which can affect gaseous ionization state, density and kinematics. The dependence of this feedback on the quasar luminosity can be studied by absorbers surrounding quasars \cite[e.g.,][]{2013JApA...34..357P,2015MNRAS.452.2553J}. We exclude quasars that the flux data at 1350\AA\ are not available by the BOSS spectra due to quasar redshifts and that are located at \zem$>2.2$, and limit absorbers with $\beta<0.03$. This results in a sample of 2866 \CIV\ NALs. We divide the resulting sample at median value of the continuum luminosity at rest frame 1350\AA\ ($L_{\rm 1350} = 10^{\rm 44.63}$ \ergs) into low- and high-luminosity quasar samples, where continuum luminosities are measured from pseudo-continuum fitting.

\begin{figure}
\centering
\includegraphics[width=7.5cm,height=6.cm]{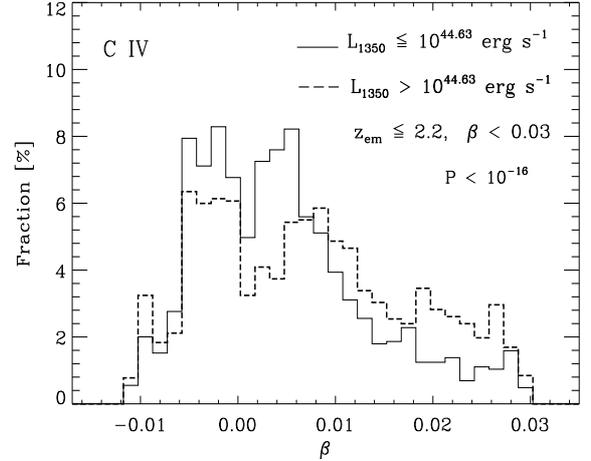}
\caption{Relative velocity distributions of absorbers with $\beta<0.03$ and \zem$\le2.2$.  The y-axis are the counts of absorbers normalized by the total numbers of absorbers contained in corresponding subsamples. The solid (dash) lines represent quasars with continuum luminosities less (greater) than corresponding median values. KS test yields a $P<10^{\rm -16}$.}
\label{fig11}
\end{figure}

\begin{figure}
\centering
\includegraphics[width=7.5cm,height=6.cm]{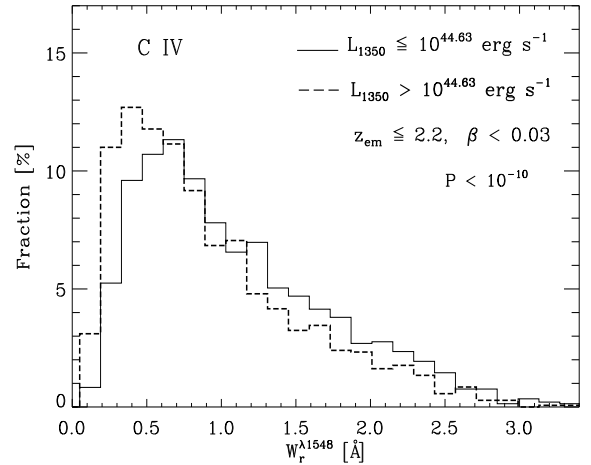}
\caption{\Wr\ distributions of absorbers with $\beta<0.03$ and \zem$\le2.2$.  The y-axis are the counts of absorbers normalized by the total numbers of absorbers contained in corresponding subsamples. KS test yields a $P<10^{\rm -10}$.}
\label{fig12}
\end{figure}

The $\beta$ and \Wr\ distributions of both low- and high-luminosity quasar samples are shown in Figures \ref{fig11} and \ref{fig12}, respectively. KS test yields a $P<10^{\rm -16}$ for the $\beta$ and a $P<10^{\rm -13}$ for the \Wr, which suggests that the ionizing radiation of quasars acts on both $\beta$ and \Wr\ distributions. We also note that the low-luminosity quasars have stronger absorption strengths.

The SDSS quasars are obtained by flux limited surveys, and consequently the high luminosity quasars can be obtained at high redshifts. Therefore, the differences of $\beta$ and \Wr\ distributions between low- and high-luminosity quasars might be the evolution result of quasar surrounding gas.
We find that both low- and high-luminosity quasars have similar median \zem. Thus, the differences reflected by Figures \ref{fig11} and \ref{fig12} are not originated in the evolution of quasar surrounding gas.

It is interesting that the high-luminosity quasars tend to show larger velocity ($\beta$) \CIV\ NALs with smaller absorption strengths, which might be related to the radiative acceleration by quasar luminosity and ionization condition of CGMs around quasars. For high-luminosity quasars, the enormous power of outflows and the energetic radiation from accretion disc could produce stronger feedback to surrounding gas of quasars. In this case, surrounding gas of high-luminosity quasars is possibly accelerated to higher velocity and photo-ionized to higher level.

\section{Summary}
\label{sect6}
Using the BOSS quasar spectra (DR9Q), combining the two early works (Papers I and II), we have detected a total number of 41,479 \CIV\ NALs with $1.4544\le$\zabs$\le4.9224$ in wavelength range between \lya\ and \CIVwave\ emission lines. The detectability tend to be complete when $S/N^{\rm \lambda1548}>4$. Ours main results are as follows.
\begin{enumerate}
  \item 92.3\% of \CIV\ NALs fall into the range of $1-\sigma_{DR}\le DR \le 2+\sigma_{DR}$, and tend to be the more saturated absorptions with the stronger absorptions that follows the curve of growth. The associated systems show significant absorptions when compared to the intervening ones.
  \item The redshift density evolution shape of \CIV\ absorbers is similar to the history of the cosmic star formation, which indicates that \CIV\ absorbers may be connected to the star formation activities within galaxies. Thus, absorption lines imprinted on the spectra of background objects are possibly a powerful tool to investigate faint and even invisible galaxies.
  \item Comparing $\beta$ distributions of \CIV\ and \MgII\ absorbers, we find clear distinctions. The $\beta$ of \CIV\ absorbers are more compact to $\beta\approx0$, and have longer extended tail. In addition, the dispersion of \CIV\ environmental absorptions is much larger than that of \MgII\ environmental absorptions. The \CIV\ has an ionization energy of 64.49 $\rm eV$, which is obviously different from that of 15.035 $\rm eV$ of the \MgII. The clearly different ionization energies of \CIV\ and \MgII\ may play an important role to the distinct $\beta$ properties between the two kinds of absorbers.
  \item For associated \CIV\ absorbers ($\beta<0.03$), when compared with the low-luminosity quasars, high-luminosity quasars tend to have larger $\beta$ and weaker absorptions.
\end{enumerate}

The line number density of absorbers depends not only on their volume density, but on physical cross-section and ionizing background of absorbing clouds. One of our future works will investigate the properties of absorbers that intersect both sightlines of a quasar pair (two close quasars on sky). This might provide a useful limit to the geometric cross-section of absorbing cloud. Deep imaging/spectroscopy observations to this kind of absorbers might also provide useful information about absorption line formation by comparing absorption strength and profile of absorption lines and characteristics of galaxy halo, such as gas distribution, star formation rate, orientations of disc and outflow, and so on.

The properties of CGMs and outflows of quasars are crucial to comprehend the quasar feedback and evolution of host galaxy. Time variability and partial coverage analysis is two efficient ways to identify absorption lines of quasar outflow and environment. Therefore, high resolution observations of the quasars with variable absorption line could provide well opportunity to investigate the properties of quasar outflow and CGM.

\vspace{6mm}The authors are very grateful to the anonymous referee for his/her critical comments and instructive suggestions, which significantly strengthened the analyses in this work. This work was supported by the National Natural Science Foundation of China (NO. 11363001, grants 11273015, 11573013 and 11133001), National Basic Research Program (973 program No. 2013CB834905), the Guangxi Natural Science Foundation (2015GXNSFBA139004), and the Guangxi University of Science and Technology research projects (No. KY2015YB289). Y.M.C. acknowledges support from the Opening Project of Key Laboratory of Computational Astrophysics, National Astronomical Observatories, Chinese Academy of Sciences.

Funding for SDSS-III has been provided by the Alfred P. Sloan Foundation, the Participating Institutions, the National Science Foundation, and the U.S. Department of Energy Office of Science. The SDSS-III web site is http://www.sdss3.org/.

SDSS-III is managed by the Astrophysical Research Consortium for the Participating Institutions of the SDSS-III Collaboration including the University of Arizona, the Brazilian Participation Group, Brookhaven National Laboratory, Carnegie Mellon University, University of Florida, the French Participation Group, the German Participation Group, Harvard University, the Instituto de Astrofisica de Canarias, the Michigan State/Notre Dame/JINA Participation Group, Johns Hopkins University, Lawrence Berkeley National Laboratory, Max Planck Institute for Astrophysics, Max Planck Institute for Extraterrestrial Physics, New Mexico State University, New York University, Ohio State University, Pennsylvania State University, University of Portsmouth, Princeton University, the Spanish Participation Group, University of Tokyo, University of Utah, Vanderbilt University, University of Virginia, University of Washington, and Yale University.

\end{document}